\documentclass[aps,pra,twocolumn,superscriptaddress]{revtex4}
\usepackage{amsmath,amssymb}
\usepackage{multirow}
\usepackage{xcolor}
\usepackage{srcltx}
\usepackage[normalem]{ulem}
\usepackage{graphicx}
\usepackage[utf8]{inputenc}

\newcommand{\ket}[1]{{|{#1}\rangle}}

\newcommand{\upb}{Integrated Quantum Optics, Universit\"at Paderborn, Warburger Strasse 100, 33098 Paderborn, Germany}

\begin{document}

\title{A two-channel, spectrally degenerate polarization entangled source on chip}

\author{Linda Sansoni}
\email{linda.sansoni@uni-paderborn.de}
\affiliation{\upb}

\author{Kai Hong Luo}
\affiliation{\upb}
\author{Christof Eigner}
\affiliation{\upb}
\author{Raimund Ricken}
\affiliation{\upb}
\author{Viktor Quiring}
\affiliation{\upb}
\author{Harald Herrmann}
\affiliation{\upb}
\author{Christine Silberhorn}
\affiliation{\upb}

\begin{abstract}

Integrated optics provides the platform for the experimental implementation of highly complex and compact circuits for quantum information applications. In this context integrated waveguide sources represent a powerful resource for the generation of quantum states of light due to their high brightness and stability. However, the confinement of the light in a single spatial mode limits the realization of multi-channel sources. Due to this challenge one of the most adopted sources in quantum information processes, i.e. a source which generates spectrally indistinguishable polarization entangled photons in two different spatial modes, has not yet been realized in a fully integrated platform. Here we overcome this limitation by suitably engineering two periodically poled waveguides and an integrated polarization splitter in lithium niobate. This source produces polarization entangled states with fidelity of $\mathcal{F}=0.973\pm0.003$ and a test of Bell's inequality results in a violation larger than 14 standard deviations. It can work both in pulsed and continuous wave regime. This device represents a new step toward the implementation of fully integrated circuits for quantum information applications.
\end{abstract}


\maketitle

In the last decade quantum photonics has played a crucial role in the development of quantum information processes. In particular the transition from bulk to integrated optics has laid the foundation for the implementation of schemes of high complexity, unachievable with previous setups.
In the framework of integrated photonic devices, a major development has been carried out in the realization of linear circuits, ranging from the implementation of single building blocks on chip \cite{poli08sci,smit09oex,sans10prl,cres11nco,heil13sre,corr14nco} to the simulation of quantum transport via quantum walks \cite{peru10sci,sans12prl,cres13npo,poul14prl}. The integration of up to a hundred optical components on the same chip has made it possible to experimentally implement the recently proposed boson sampling problem \cite{aaro13toc,till13npo,cres13npo2,spri13sci,broo13sci,spag14npo,bent15sad} opening the way to the experimental investigation of classically hard to solve problems.
Although the realization of integrated linear circuits has improved significantly, the conversion toward a fully integrated platform for quantum information requires the development and the combination of sources of quantum light and eventually detectors on chip. Concerning sources on chip, parametric down conversion (PDC) in periodically poled nonlinear waveguides is a promising approach \cite{mart10njp,herr13oex,krap13njp,kais13lpl} due to high brightness and intrinsic stability. However the confinement of light in a single channel makes it challenging to emit single photons into different spatial modes without the need of postselection.
This is the case of one of the most adopted sources in quantum information experiments, i.e. a source of spectrally degenerate polarization entangled states \cite{kwia95prl,fedr07oex}. While the emission into two different spatial modes is intrinsic in bulk crystals, in integrated waveguides this still remains a challenge and the main reason why a fully on chip source of polarization entangled states was still missing.
Indeed nonlinear waveguides have been adopted to realize sources of polarization entangled states, but the entanglement was either generated outside the chip with a probabilistic splitting on different spatial modes \cite{mart10njp,kais13lpl}, or the pairs were split deterministically by frequency, leading to spectrally non-degenerate photons \cite{herr13oex}. While for some applications this strategy can be useful, degeneracy over all degrees of freedom is a fundamental requirement when dealing with quantum tasks where completely indistinguishable particles are needed.  Recently it has been demonstrated that parametric down conversion in a coupled structure allows for the generation of path entangled states \cite{krus15pra,setz16lpr}. However the adoption of this concept to deterministically address different spatial modes and create entanglement in other degrees of freedom is not straightforward.

Here we overcome these limitations by exploiting the features of titanium indiffused waveguides on lithium niobate (LN) substrates and realize a fully on chip source of polarization entangled states at degenerate wavelengths in the telecom regime. These waveguides can indeed guide both polarizations and present high $\chi^{(2)}$ nonlinearity.
\begin{figure}[ht!!]
\centering
\includegraphics[width=0.95\columnwidth]{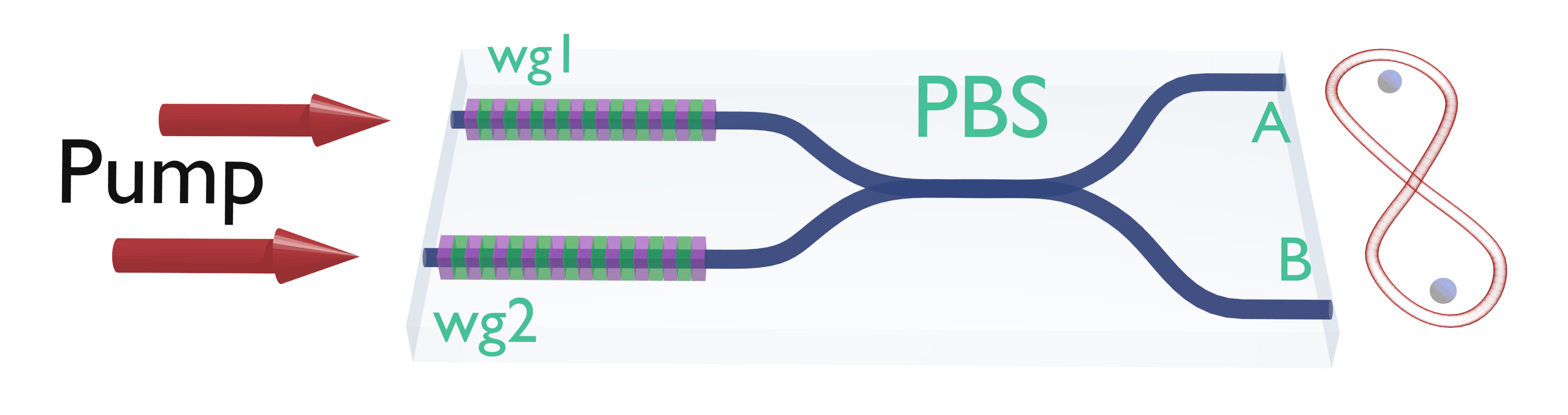}
\caption{Chip design: two periodically poled waveguides produce orthogonally polarized photons via type II parametric down conversion. The pairs are split on an integrated polarizing beam splitter (PBS) such that the output state is $\ket{H_A,V_B}$ ($\ket{V_A,H_B}$) when the pair is generated in waveguide wg1 (wg2). The superposition of these two contributions results in the desired entangled state.}
\label{fig:sourcescheme}
\end{figure}
In order to achieve polarization entanglement, we implemented a scheme as shown in Fig. \ref{fig:sourcescheme}. Here, two periodically poled nonlinear waveguides designed to generate orthogonally polarized photons via type II parametric down conversion are connected to a zero-gap directional coupler acting as a polarizing beam splitter (PBS). When a pair is generated in waveguide 1 (wg1) the output state is $\ket{H_A,V_B}$, where $H_i$ ($V_i$) represents a horizontal (vertical) polarized photon in the output mode $i=A,B$; while when the pair is generated in waveguide 2 (wg2) the resulting state is $\ket{V_A,H_B}$. By enabling PDC in both waveguides the superposition of these two contributions yields to the desired entangled state $\frac{1}{\sqrt{2}}[\ket{H_A,V_B}+e^{i\phi}\ket{V_A,H_B}]$. The phase $\phi$ is now defined by the phase between the two pump beams.\\
We can describe the action of our device through the transformation from input to output modes of the integrated PBS as a function of transmissivity and reflectivity:
 \begin{equation}
 \left\{
 \begin{matrix}
 a^{\dagger}_{1,H}=\sqrt{T_{H}}\ a^{\dagger}_{A,H}-\sqrt{R_{H}}\ a^{\dagger}_{B,H}\\
 a^{\dagger}_{2,H}=\sqrt{R_{H}}\ a^{\dagger}_{A,H}+\sqrt{T_{H}}\ a^{\dagger}_{B,H}\\
 a^{\dagger}_{1,V}=\sqrt{T_{V}}\ a^{\dagger}_{A,V}-\sqrt{R_{V}}\ a^{\dagger}_{B,V}\\
 a^{\dagger}_{2,V}=\sqrt{R_{V}}\ a^{\dagger}_{A,V}+\sqrt{T_{V}}\ a^{\dagger}_{B,V}
 \end{matrix}
 \right.
 \label{eq:PBS}
 \end{equation}
Here $a^{\dagger}_{k,J}$ is a creation operator acting on a photon in mode $k$ with polarization $J$ and $T_{J}=1-R_{J}$ are the transmissivity and reflectivity of a photon with polarization $J$ in either modes.
The state generated in the two periodically poled waveguides when both are pumped is $\ket{\psi_{in}}=\frac{1}{\sqrt{2}}[a^{\dagger}_{1,H}a^{\dagger}_{1,V}+e^{i\phi}a^{\dagger}_{2,H}a^{\dagger}_{2,V}]\ket{0}$. The creation operators evolve according to transformations (\ref{eq:PBS}) and then the output state will read
\begin{equation}
\begin{split}
	\ket{\psi_{out}} =\frac{1}{\sqrt{2}}
	[&(\sqrt{T_{V}T_{H}}+e^{i\phi}\sqrt{R_{V}R_{H}})a^{\dagger}_{A,H}a^{\dagger}_{A,V}\\
	+& (\sqrt{R_{V}T_{H}}-e^{i\phi}\sqrt{T_{V}R_{H}})a^{\dagger}_{A,H}a^{\dagger}_{B,V}\\
	 +& (\sqrt{T_{V}R_{H}}-e^{i\phi}\sqrt{R_{V}T_{H}})a^{\dagger}_{A,V}a^{\dagger}_{B,H}\\
	 +& (\sqrt{R_{V}R_{H}}+e^{i\phi}\sqrt{T_{V}T_{H}})a^{\dagger}_{B,H}a^{\dagger}_{B,V}]\ket{0}
\end{split}
\end{equation}
In the ideal case $R_{V}=T_{H}=1$ and the device produces the expected entangled state $\ket{\psi}_{ent}=\frac{1}{\sqrt{2}}[a^{\dagger}_{A,H}a^{\dagger}_{B,V}-e^{i\phi}a^{\dagger}_{B,H}a^{\dagger}_{A,V}]\ket{0}$. Deviations from these ideal values lead to a non vanishing probability of having two photons in the same output (for a detailed description see Supplementary Material).

This design has two main challenges: the fabrication of two identical periodically poled waveguides such that the PDC light produced in both of them has the same spectral properties and the realization of an integrated PBS with polarization dependent splitting ratios as close as possible to the ideal ones.\\
Periodically poled titanium indiffused waveguides on lithium niobate substrates are the optimal candidate to realize this design with the required quality: they guide both polarizations, present very low losses and high $\chi^{(2)}$ nonlinearity. Moreover, we have achieved a control over the fabrication technique such that our source has excellent performances and higher brightness compared to bulk sources due to the combination of high $\chi^{(2)}$ nonlinearity of lithium niobate and waveguide confinement.

\begin{figure}[ht]
\centering
\includegraphics[width=0.9\columnwidth]{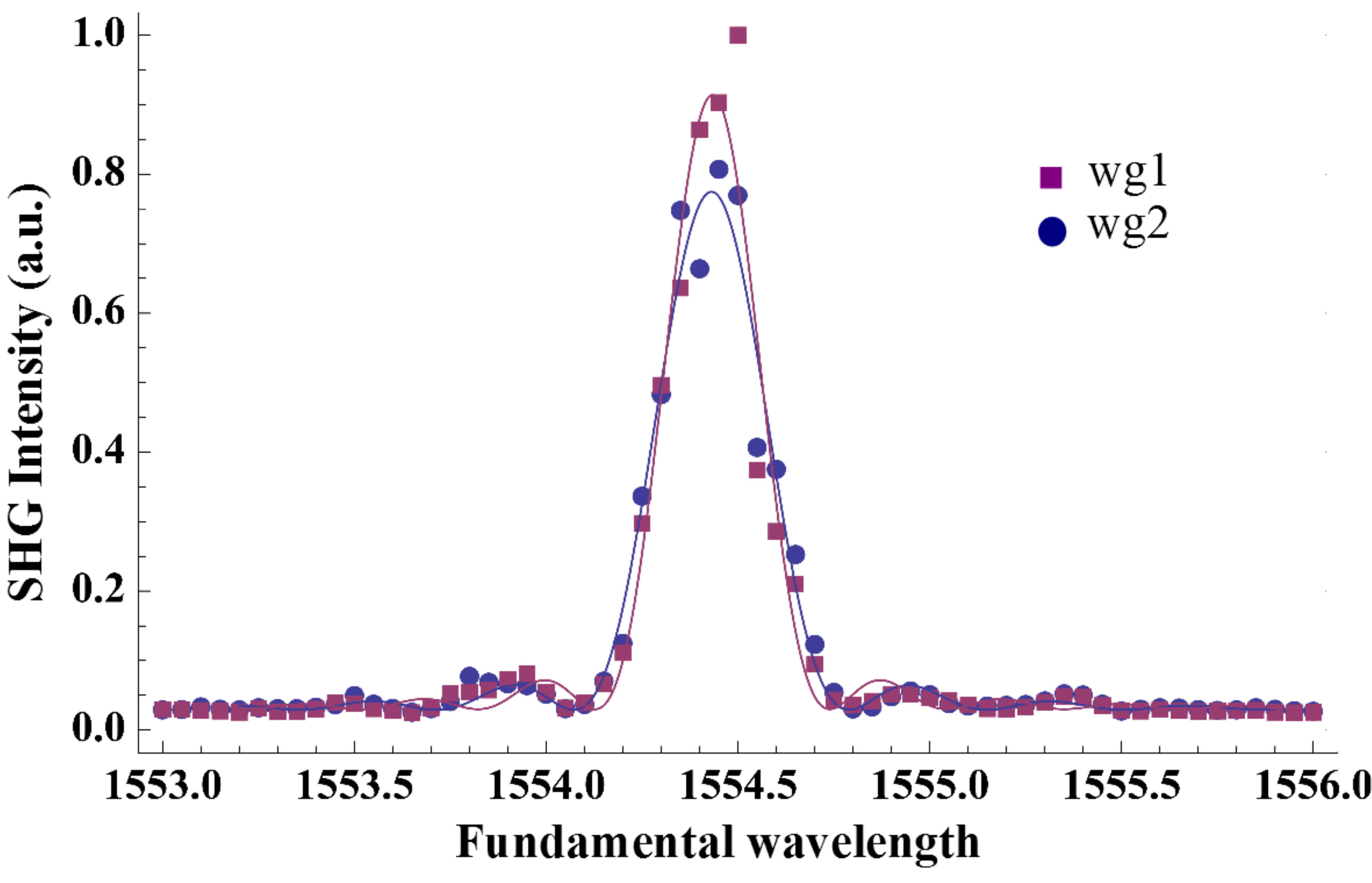}
\caption{SHG measurements: purple squares correspond to SHG process in wg1 and blue dots to SHG in wg2. Continuous lines are fits. It is clear that the two processes overlap. Error bars are smaller than the size of the markers.}
\label{fig:SHG}
\end{figure}
We fabricated a chip with the following parameters: the periodically poled waveguides have a length of $24\ mm$ with poling period of $\Lambda=9.08\ \mu m$. Spectrally degenerate pairs at $1554.44\ nm$ are generated when the sample is at a temperature of $T=50^{\circ}C$ via type II PDC.
The polarization dependent splitting is achieved through a zero-gap coupler (see further details in the Supplementary Material). We optimized the design of such a coupler to achieve optimal splitting ratios for horizontal and vertical polarized light. The transmission for H and reflectivity for V are respectively $T_H=0.996\pm0.002$ and $R_V=0.968\pm0.003$. This corresponds to a probability of generating the singlet state of $\mathcal{P}_{\ket{\psi^-}}=0.942$ (See Supplementary Material for detailed calculations).
The total length of our chip is $L=49\ mm$. The output facet is coated to highly reflect the pump wavelength around $775\ nm$ and to highly transmit the PDC pairs at telecom wavelengths. We achieve a suppression of $\approx22\ dB$ for the pump and a transmissivity $\geq 98\%$ for PDC light. Our device presents very low losses, as expected for LN, the average value being $0.03\ dB/cm$.
\begin{figure*}[ht!!]
\centering
\includegraphics[width=0.9\textwidth]{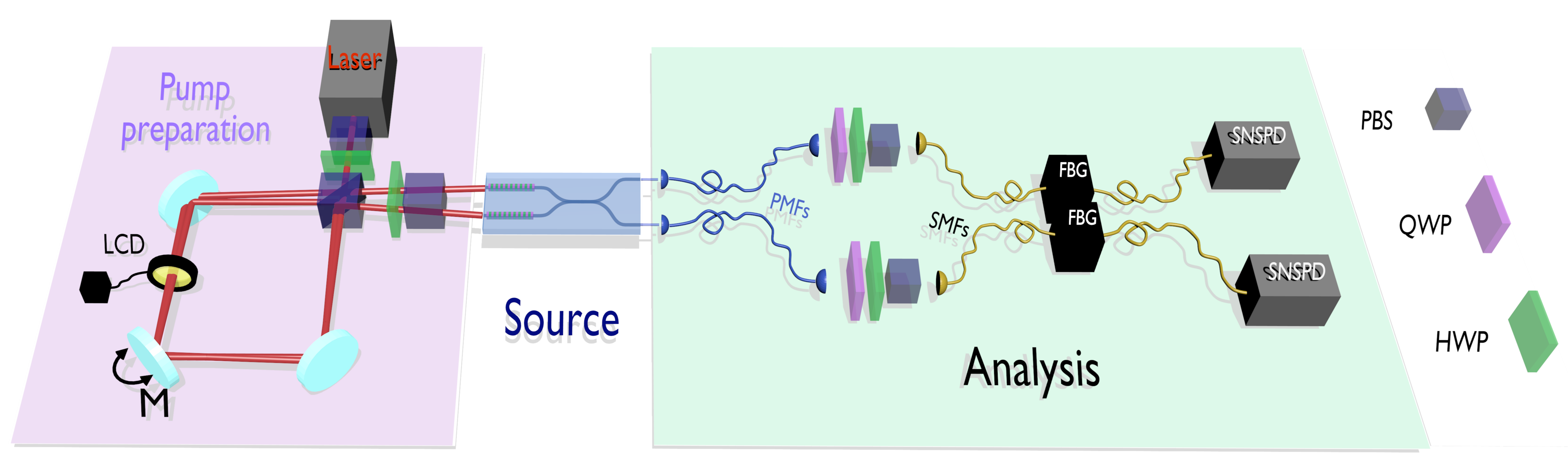}
\caption{Experimental setup for quantum measurements: the two pump beams are created within a Sagnac loop and displaced by rotating one mirror of the loop (M). The phase between the two beams is set via a liquid crystal device (LCD). After the chip the entangled state is measured with a standard polarization analysis setup: half waveplate (HWP), quarter waveplate (QWP) and a polarizing beam splitter (PBS). The photons are detected with superconducting nanowire single photon detector (SNSPD). Temporal delay is compensated with polarization maintaining fibers (PMF) and fiber Bragg grating filters (FBG) can be possibly inserted to filter the spectrum of the emitted photons.}
\label{fig:setup}
\end{figure*}

We tested the nonlinear properties of the two periodically poled waveguides via second harmonic generation (SHG) in order to find the correct wavelength for a spectrally degenerate process. A continuous wave (CW) bright field at telecom wavelength has been injected in each waveguide singularly and photons at half of the wavelength have been detected with a PIN-diode. By varying the wavelength of the bright field the curves in Fig. \ref{fig:SHG} are recorded. The full width at half maximum (FWHM) of the two curves are $\Delta\lambda_{wg1}=(0.306\pm0.009)nm$ and $\Delta\lambda_{wg2}=(0.359\pm0.008)nm$ for waveguide 1 and 2 respectively and both the processes are phase matched at $\lambda=(1554.44\pm0.01)nm$. The slight difference in the FWHMs is due to a slightly different effective poling length of the waveguide, arising from fabrication imperfections. 
Nevertheless the overlap between the two curves is $\mathcal{O}\approx0.97$, i.e. the two poled waveguides present almost the same nonlinear properties, which is essential for the correct functioning of the device.
The peak wavelength of the SHG curve also tells us the wavelength combination to phase match the corresponding degenerate PDC process. It corresponds to $\lambda_p=777.22\ nm\rightarrow\lambda_s=\lambda_i=1554.44\ nm $ (subscripts $p,s,i$ stay for pump, signal and idler respectively). This is the process we will exploit to produce spectrally degenerate polarization entangled states.

A fundamental requirement is now that the photons have no temporal distinguishability, otherwise it will wash out the entanglement. Lithium niobate is a birefringent material, thus horizontal and vertical polarized photons experience different group velocities when traveling through the waveguides. We can consider that the photons are generated at the center of the poled region of length $L_p$, this yields a delay between H and V photons which is proportional to an effective crystal length $L_{eff}=L-L_p/2$:
\begin{equation}
\Delta\tau=\frac{(n_{gH}-n_{gV})L_{eff}}{c}
\end{equation}
where $n_{gJ}$ is the group refractive index for $J=H,V$ polarization in LN and $c$ the speed of light in vacuum. 
In order to compensate for this temporal walk off we adopt polarization maintaining fibers (PMFs) and let the fast polarized photon travel through the slow axis of the fiber and vice versa. The length of the fibers $L_{f}$ is chosen such that the delay between H and V in the fiber is
\begin{equation}
\Delta\tau_{f}=\frac{B\ L_{f}}{c}=\Delta\tau
\end{equation}
$B$ being the birefringence of the PMF.
In our experiment the length of our chip is $L=49\ mm$ and the periodic poled waveguides occupy $L_p=24\ mm$ of it. The effective length is then $L_{eff}=L-L_p/2=37\ mm$, which introduces a delay $\Delta\tau=9.31\ ps$. The length of polarization maintaining fibers is chosen accordingly to be $L_f=6.95\ m$. Let us note that the PMFs can be directly glued to our chip and be considered part of the integrated source itself. However, for some applications, such a temporal compensation can be applied after the total evolution of the two-photon states through a linear network; in this case the PMFs can be considered part of the detection scheme, implying that the presence of the PMFs does not prevent the possibility of directly interfacing this LN chip with other integrated circuits.

A second source of distinguishability that we should prevent or compensate is a possible non-perfect overlap between the PDC spectra of H and V photons. This is directly related to symmetry of the joint spectral amplitude (JSA) of the PDC light produced in each waveguide (see Supplementary Material for more details). For a type II process, where orthogonally polarized pairs are emitted, this condition is true only when the pump bandwidth is small enough to ensure a JSA oriented at $-45^{\circ}$ in the plane of signal and idler wavelengths. This case correspond to a process driven by CW pump light. On the contrary the use of pulses results in an asymmetry of the JSA and a consequent difference in the bandwidths of the spectra for H and V photons. In this case spectral filters have to be introduced. We pumped our source with both CW and picosecond pulses. In the case of pulsed light we adopted fiber Bragg grating (FBG) filters which, as in the case of PMFs, can be considered part of the source or of the detection scheme.
The setup for quantum measurements is shown in Fig. \ref{fig:setup}: it can be divided in three sections: the pump preparation setup, the source itself and the analysis setup. The pump preparation setup consists of a loop: here a diagonally polarized beam is split on a bulk PBS and fed into a Sagnac loop. The two beams are recombined on the PBS, however, no interference occurs because a displacement between them is introduced by tilting one mirror ($M$) of the loop. This displacement is imaged to the separation between the two input waveguides of the chip through a set of lenses (not shown in the picture). Such a configuration ensures a common path geometry for the two pump beams such that their relative phase remains constant in time. The desired phase $\phi$ between them is then set via a liquid crystal device (LCD). After the loop a half waveplate and another PBS are used to set the polarization state of both beams (horizontal). Let us note that by properly rotating the waveplates before and after the Sagnac loop we are able to pump only waveguide 1 or waveguide 2 and generate separable states, or both at the same time and generate polarization entanglement.
The second part of the setup is the source itself: the LN chip which generates the polarization entangled state.
The last part is the analysis setup: PMFs are used for temporal delay compensation. In our scheme we chose to use a free space coupling but it is also possible to directly interface a fiber array of PMFs to the output of the chip, as the waveguide separation at the output matches the standard separation between fibers in fiber arrays ($s=127\ \mu m$). In the case of pulsed pump beams FBG filters with bandwidth $\Delta\lambda=0.25\ nm$  are inserted to apply spectral filtering. Standard polarization analysis components and superconducting nanowires single photon detectors (SNSPD, Quantum Opus) are used to characterize the state. Coincidence counts between the two outputs are recorded.
\begin{figure}[ht!!]
\centering
\includegraphics[width=0.9\columnwidth]{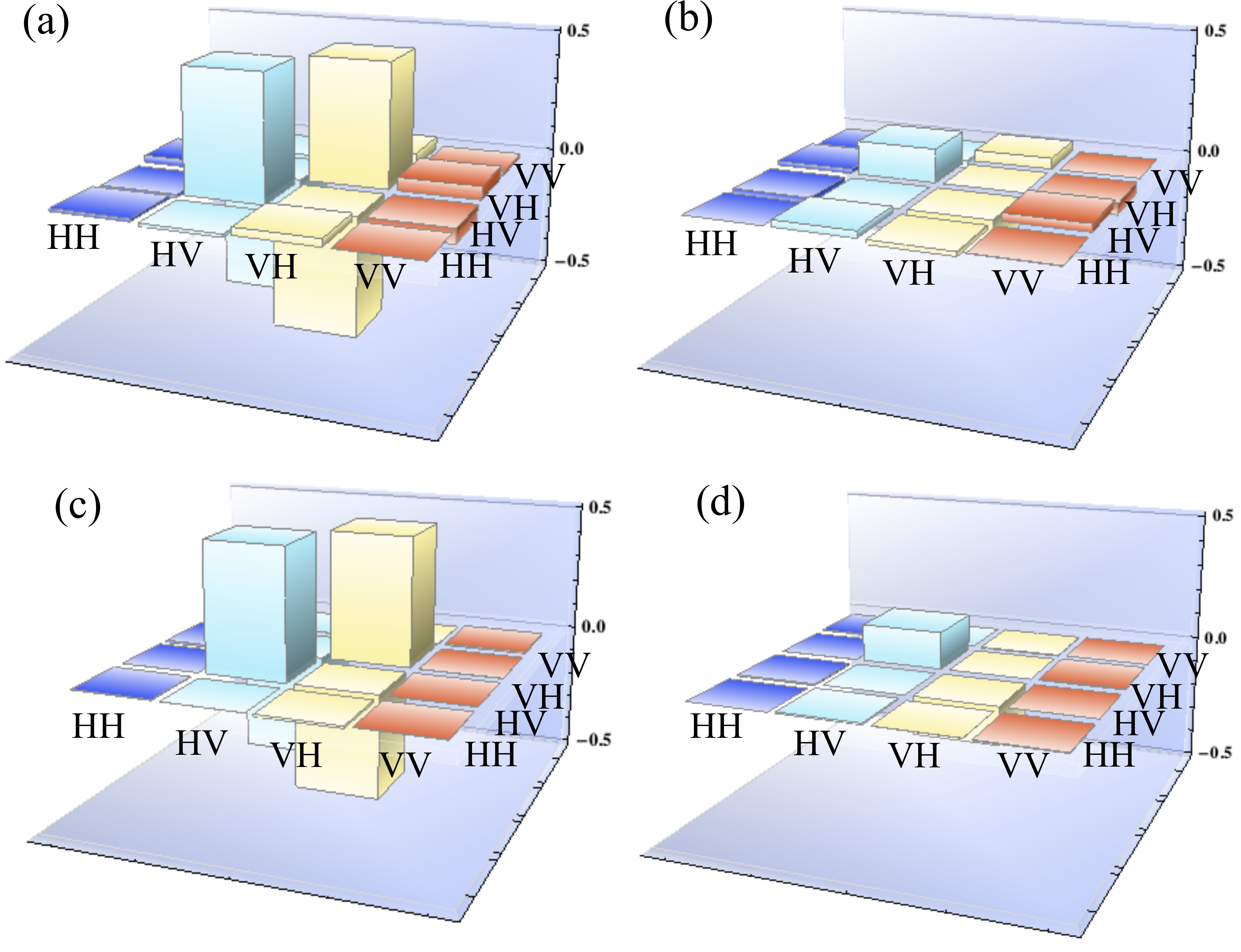}
\caption{Real (a) and imaginary part (b) of the experimental density matrix of the state produced by our source when pumped with pulsed light. Real (c) and imaginary (d) part of experimental density matrix obtained with CW pump light and no filters. Small contributions in the imaginary part are due to non perfect alignment of the axes of the PMFs with the polarization analysis setup.}
\label{fig:qst}
\end{figure}
By measuring single counts and coincidences between the two SNSPDs we evaluated the high brightness of our source to be $\mathcal{B}=(4.84\pm0.09)\cdot10^6 \frac{\text{pairs}}{s\ nm\ mW}$. Then, in order to characterize our quantum state, we performed a quantum state tomography \cite{jame01pra} to reconstruct the density matrix describing the produced state.
We set the pump wavelength at $777.22\ nm$ to ensure spectral degeneracy and measured all the possible polarization states projections. Real and imaginary parts of the experimental density matrix are reported in Fig. \ref{fig:qst} \textit{a,b} for pulsed and in Fig. \ref{fig:qst} \textit{c,d} for CW pump light. For pulsed pump light the fidelity (overlap) between the measured state and the singlet state of the Bell basis is $\mathcal{F}_{ps}=0.973\pm0.003$. This value is mainly affected by the imperfect splitting ratio of the PBS on chip, residual temporal mismatch and possible slight mismatch of the central peak of the adopted FBGs.
We then performed a Bell test according to the scheme of Clause, Horne, Shimony and Holt (CHSH) \cite{clau69prl} and obtained a value of the $S$-parameter of $S_{ps}=2.694\pm0.046$ corresponding to a violation of more than 14 standard deviations.
We performed the same measurements with a CW pump and no filters and obtained a fidelity of $\mathcal{F}_{CW}=0.941\pm0.002$ and S-parameter $S_{CW}=2.597\pm0.027$. This fidelity is lower compared to the pulsed pump case because of slightly different properties between the two poled waveguides (shown already in the SHG measurements). However this can be improved in the fabrication process.

These results show that our fully integrated source of polarization entangled states has high quality performances that make it the optimal candidate to replace bulk sources commonly adopted in quantum information setups. In this device, the high brightness, low losses and the ability of guiding both polarizations, in combination with a smart waveguide design allow us to achieve a considerable step forward in the realization of integrated multi-channel sources which are one fundamental ingredient of a fully integrated platform for quantum information processes.

\noindent\textbf{Funding Information}
This work has received funding from the European Union's Horizon 2020 research and innovation program under the QUCHIP project GA no. 641039 and from the Deutsche Forschungsgemeinschaft (DFG) via the Gottfried Wilhelm Leibniz-Preis.


\begin{thebibliography}{10}
\newcommand{\enquote}[1]{``#1''}

\bibitem{poli08sci}
A.~Politi, M.~J. Cryan, J.~G. Rarity, S.~Yu, and J.~L. O'Brien,
  \enquote{Silica-on-silicon waveguide quantum circuits,} Science \textbf{320},
  646--649 (2008).

\bibitem{smit09oex}
B.~J. Smith, D.~Kundys, N.~Thomas-Peter, P.~G.~R. Smith, and I.~A.Walmsley,
  \enquote{Phase-controlled integrated photonic quantum circuits,} Optics
  Express \textbf{17}, 13516--13525 (2009).

\bibitem{sans10prl}
L.~Sansoni, F.~Sciarrino, G.~Vallone, P.~Mataloni, A.~Crespi, R.~Ramponi, and
  R.~Osellame, \enquote{Polarization entangled state measurement on a chip,}
  Physical Review Letters \textbf{105}, 200503 (2010).

\bibitem{cres11nco}
A.~Crespi, R.~Ramponi, R.~Osellame, L.~Sansoni, I.~Bongioanni, F.~Sciarrino,
  G.~Vallone, and P.~Mataloni, \enquote{Integrated photonic quantum gates for
  polarization qubits,} Nature Communications \textbf{2}, 566 (2011).

\bibitem{heil13sre}
R.~Heilmann, M.~Gr\"afe, S.~Nolte, and A.~Szameit, \enquote{Arbitrary photonic
  wave plate operations on chip: Realizing {Hadamard, Pauli-X}, and rotation
  gates for polarisation qubits,} Scientific Report \textbf{4}, 4118 (2013).

\bibitem{corr14nco}
G.~Corrielli, A.~Crespi, R.~Geremia, R.~Ramponi, L.~Sansoni, A.~Santinelli,
  P.~Mataloni, F.~Sciarrino, and R.~Osellame, \enquote{Rotated waveplates in
  integrated waveguide optics,} Nature Communications \textbf{5}, 4249 (2014).

\bibitem{peru10sci}
A.~Peruzzo, M.~Lobino, J.~C.~F. Matthews, N.~Matsuda, A.~Politi, K.~Poulios,
  X.-Q. Zhou, Y.~Lahini, N.~Ismail, K.~Worhoff, Y.~Bromberg, Y.~Silberberg,
  M.~G. Thompson, and J.~L. O'Brien, \enquote{Quantum walks of correlated
  photons,} Science \textbf{329}, 1500--1503 (2010).

\bibitem{sans12prl}
L.~Sansoni, F.~Sciarrino, G.~Vallone, P.~Mataloni, A.~Crespi, R.~Ramponi, and
  R.~Osellame, \enquote{Two-particle bosonic-fermionic quantum walk via
  integrated photonics,} Physical Review Letters \textbf{108}, 010502 (2012).

\bibitem{cres13npo}
A.~Crespi, R.~Osellame, R.~Ramponi, V.~Giovannetti, R.~Fazio, L.~Sansoni, F.~D.
  Nicola, F.~Sciarrino, and P.~Mataloni, \enquote{Anderson localization of
  entangled photons in an integrated quantum walk,} Nature Photonics
  \textbf{7}, 322--328 (2013).

\bibitem{poul14prl}
K.~Poulios, R.~Keil, D.~Fry, J.~D. Meinecke, J.~C. Matthews, A.~Politi,
  M.~Lobino, M.~Gr\"afe, M.~Heinrich, S.~Nolte, A.~Szameit, and J.~L. O'Brien,
  \enquote{Quantum walks of correlated photon pairs in two-dimensional
  waveguide arrays,} Physical Review Letters \textbf{112}, 143604 (2014).

\bibitem{aaro13toc}
S.~Aaronson and A.~Arkhipov, \enquote{The computational complexity of linear
  optics,} Theory of Computing \textbf{9}, 143--252 (2013).

\bibitem{till13npo}
M.~Tillmann, B.~Daki{\`c}, R.~Heilmann, S.~Nolte, A.~Szameit, and P.~Walther,
  \enquote{Experimental boson sampling,} Nature Photonics \textbf{7}, 540
  (2013).

\bibitem{cres13npo2}
A.~Crespi, R.~Osellame, R.~Ramponi, D.~J. Brod, E.~F.~G. ao, N.~Spagnolo,
  C.~Vitelli, E.~Maiorino, P.~Mataloni, and F.~Sciarrino, \enquote{Integrated
  multimode interferometers with arbitrary designs for photonic boson
  sampling,} Nature Photonics \textbf{7}, 545--549 (2013).

\bibitem{spri13sci}
J.~B. Spring, B.~J. Metcalf, P.~C. Humphreys, W.~S. Kolthammer, X.~Jin,
  M.~Barbieri, A.~Datta, N.~Thomas-Peter, N.~K. Langford, D.~Kundys, J.~C.
  Gates, B.~J. Smith, P.~G.~R. Smith, and I.~A. Walmsley, \enquote{Boson
  sampling on a photonic chip,} Science \textbf{339} (2013).

\bibitem{broo13sci}
M.~A. Broome, A.~Fedrizzi, S.~Rahimi-Keshari, J.~Dove, S.~Aaronson, T.~C.
  Ralph, and A.~G. White, \enquote{Photonic boson sampling in a tunable
  circuit,} Science \textbf{339} (2013).

\bibitem{spag14npo}
N.~Spagnolo, C.~Vitelli, M.~Bentivegna, D.~J. Brod, A.~Crespi, F.~Flamini,
  S.~Giacomini, G.~Milani, R.~Ramponi, P.~Mataloni, R.~Osellame, E.~F. Galvao,
  and F.~Sciarrino, \enquote{Experimental validation of photonic boson
  sampling,} Nat Photon \textbf{8}, 615--620 (2014).

\bibitem{bent15sad}
M.~Bentivegna, N.~Spagnolo, C.~Vitelli, F.~Flamini, N.~Viggianiello,
  L.~Latmiral, P.~Mataloni, D.~J. Brod, E.~F. Galvao, A.~Crespi, R.~Ramponi,
  R.~Osellame, and F.~Sciarrino, \enquote{Experimental scattershot boson
  sampling,} Science Advances \textbf{1}, e1400255 (2015).

\bibitem{mart10njp}
A.~Martin, A.~Issautier, H.~Herrmann, W.~Sohler, D.~Ostrowsky, O.~Alibart, and
  S.~Tanzilli, \enquote{A polarization entangled photon-pair source based on a
  {Type-II PPLN} waveguide emitting at a telecom wavelength,} New Journal of
  Physics \textbf{12}, 103005 (2010).

\bibitem{herr13oex}
H.~Herrmann, X.~Yang, A.~Thomas, A.~Poppe, W.~Sohler, and C.~Silberhorn,
  \enquote{Post-selection free, integrated optical source of non-degenerate,
  polarization entangled photon pairs,} Opt. Express \textbf{21}, 27981--27991
  (2013).

\bibitem{krap13njp}
S.~Krapick, H.~Herrmann, V.~Quiring, B.~Brecht, H.~Suche, and C.~Silberhorn,
  \enquote{An efficient integrated two-color source for heralded single
  photons,} New Journal of Physics \textbf{15}, 033010 (2013).

\bibitem{kais13lpl}
F.~Kaiser, A.~Issautier, L.~A. Ngah, O.~Alibart, A.~Martin, and S.~Tanzilli,
  \enquote{A versatile source of polarisation entangled photons for quantum
  network applications,} Laser Physics Letters \textbf{10}, 045202 (2013).

\bibitem{kwia95prl}
P.~Kwiat, K.~Mattle, H.~Weinfurter, and A.~Zeilinger, \enquote{New
  high-intensity source of polarization entangled photon pairs,} Physical
  Review Letters \textbf{75}, 4337 (1995).

\bibitem{fedr07oex}
A.~Fedrizzi, T.~Herbst, A.~Poppe, T.~Jennewein, and A.~Zeilinger, \enquote{A
  wavelength-tunable fiber-coupled source of narrowband entangled photons,}
  Optics Express \textbf{15}, 15377 (2007).

\bibitem{krus15pra}
R.~Kruse, L.~Sansoni, S.~Brauner, R.~Ricken, C.~S. Hamilton, I.~Jex, and
  C.~Silberhorn, \enquote{Dual-path source engineering in integrated quantum
  optics,} Phys. Rev. A \textbf{92}, 053841 (2015).

\bibitem{setz16lpr}
F.~Setzpfandt, A.~S. Solntsev, J.~Titchener, C.~W. Wu, C.~Xiong, R.~Schiek,
  T.~Pertsch, D.~N. Neshev, and A.~A. Sukhorukov, \enquote{Tunable generation
  of entangled photons in a nonlinear directional coupler,} Laser \& Photonics
  Reviews \textbf{10}, 131--136 (2016).

\bibitem{jame01pra}
D.~F.~V. James, P.~G. Kwiat, W.~J. Munro, and A.~G. White, \enquote{Measurement
  of qubits,} Physical Review A \textbf{64}, 052312 (2001).

\bibitem{clau69prl}
J.~Clauser, M.~Horne, A.~Shimony, and R.~Holt, \enquote{Proposed experiment to
  test local hidden-variable theories,} Physical Review Letters \textbf{23},
  880 (1969).

\end{thebibliography}

\newpage

\section*{SUPPLEMENTARY MATERIAL}

\section*{Theoretical description of state generation}
\begin{figure}[hb!!]
\centering
\includegraphics[width=0.8\columnwidth]{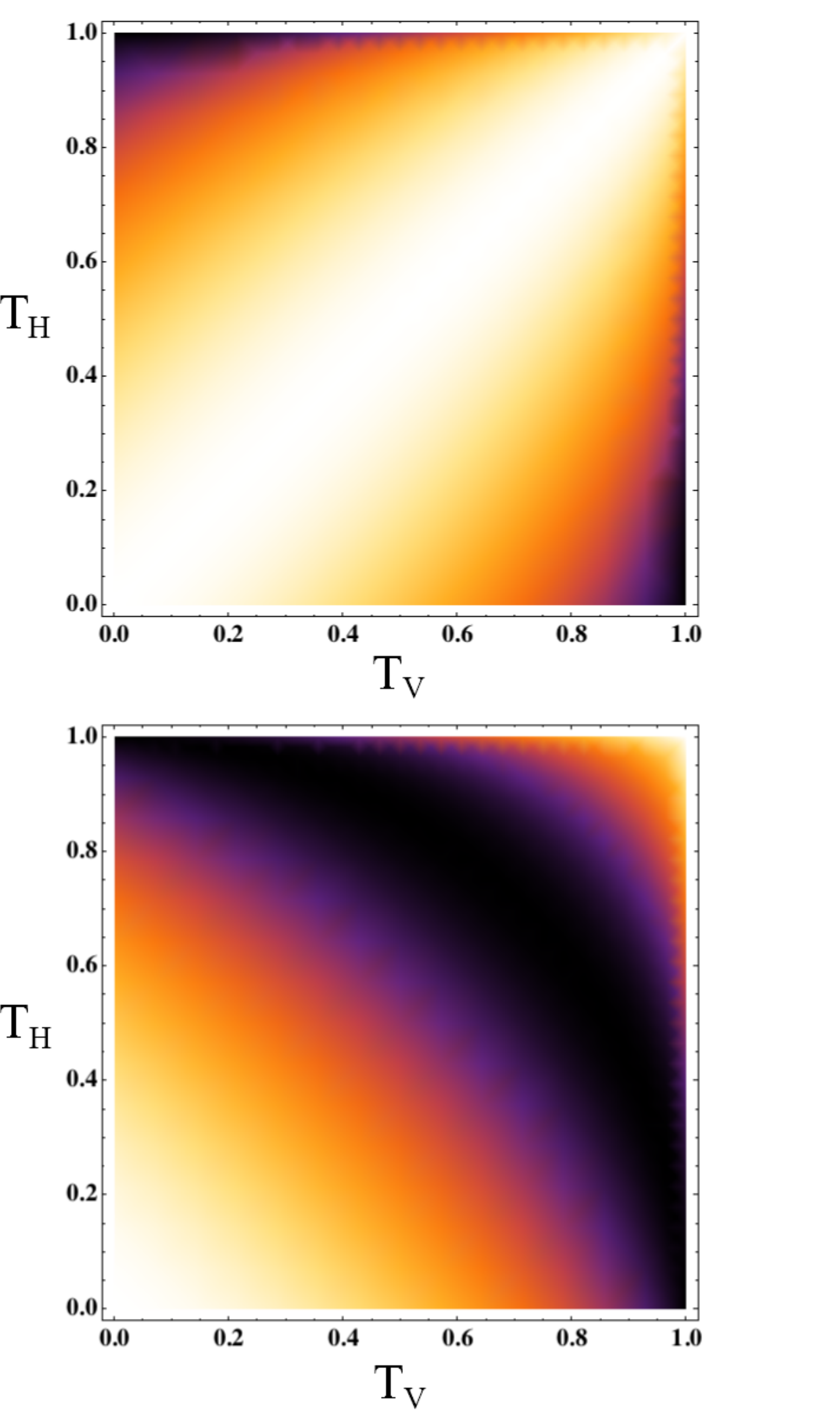}
\caption{Probability of generating $\ket{\psi^-}$ (top) and $\ket{\psi^+}$ (bottom) as a function of transmission for H and V polarizations. Ideal values are $T_H=1$ and $T_V=0$ which correspond to a probability $\mathcal{P}_{\ket{\psi^{\pm}}}=1$. The probability of generating $\ket{\psi^-}$ is more affected by deviations of the transmissivities from ideal values.}
\label{fig:prob}
\end{figure}

As described in the main text, the state generated by our device can be described through the transformation induced by the integrated polarization dependent splitter on the photons generated into the periodically poled waveguides.
The initial state is
\begin{equation}
	\ket{\psi_{in}}=\frac{1}{\sqrt{C_1^2+C_2^2}}[[C_1\ a^{\dagger}_{1,H}a^{\dagger}_{1,V}+C_2\ e^{i\phi}a^{\dagger}_{2,H}a^{\dagger}_{2,V}]\ket{0}
	\label{eq:instate}
\end{equation}
where  $a^{\dagger}_{k,J}$ is a creation operator acting on a photon in mode $k$ with polarization $J$, and $C_{1,2}$ weight the contribution of each source.  Let us note that if the efficiencies of the two periodically poled sections are different we can achieve a maximally entangled state, i.e. $C_1=C_2$, by adjusting the pump power in order to compensate for the different efficiencies. We assume in the following this condition. We recall here the transformation induced by the PBS:
 \begin{equation}
 \left\{
 \begin{matrix}
 a^{\dagger}_{1,H}=\sqrt{T_{H}}\ a^{\dagger}_{A,H}-\sqrt{R_{H}}\ a^{\dagger}_{B,H}\\
 a^{\dagger}_{2,H}=\sqrt{R_{H}}\ a^{\dagger}_{A,H}+\sqrt{T_{H}}\ a^{\dagger}_{B,H}\\
 a^{\dagger}_{1,V}=\sqrt{T_{V}}\ a^{\dagger}_{A,V}-\sqrt{R_{V}}\ a^{\dagger}_{B,V}\\
 a^{\dagger}_{2,V}=\sqrt{R_{V}}\ a^{\dagger}_{A,V}+\sqrt{T_{V}}\ a^{\dagger}_{B,V}
 \end{matrix}
 \right.
 \label{eq:PBS}
 \end{equation}
Here $T_{J}=1-R_{J}$ are the transmissivity and reflectivity of a photon with polarization $J$ in either modes.
After applying these transformations to the input state (\ref{eq:instate}) the output state will read:
\begin{equation}
\begin{split}
	\ket{\psi_{out}} =\frac{1}{\sqrt{2}}
	[&(\sqrt{T_{V}T_{H}}+e^{i\phi}\sqrt{R_{V}R_{H}})a^{\dagger}_{A,H}a^{\dagger}_{A,V}\\
	+& (\sqrt{R_{V}T_{H}}-e^{i\phi}\sqrt{T_{V}R_{H}})a^{\dagger}_{A,H}a^{\dagger}_{B,V}\\
	 +& (\sqrt{T_{V}R_{H}}-e^{i\phi}\sqrt{R_{V}T_{H}})a^{\dagger}_{A,V}a^{\dagger}_{B,H}\\
	 +& (\sqrt{R_{V}R_{H}}+e^{i\phi}\sqrt{T_{V}T_{H}})a^{\dagger}_{B,H}a^{\dagger}_{B,V}]\ket{0}
\end{split}
\end{equation}
Now it is interesting to note that the coefficients of the four terms depend on the phase $\phi$ and this dependency vanishes only for ideal parameters $T_H=R_V=1$.
Deviations from these ideal values increase the probability of having two photons in the same output and consequently decrease the probability of generating the desired entangled state.
We can define the probability of generating the entangled state as follows:
\begin{equation}
\begin{split}
	\mathcal{P}_{\ket{\psi^{\pm}}}(T_H,T_V,\phi)=&|\langle \psi^{\pm}|\psi_{out}\rangle|^2\\
	=&\frac{1}{2}(|\sqrt{R_{V}T_{H}}-e^{i\phi}\sqrt{T_{V}R_{H}}|^2\\
	&+|\sqrt{R_{V}R_{H}}+e^{i\phi}\sqrt{T_{V}T_{H}}|^2)
\end{split}
\end{equation}
This probability is plot in Fig. \ref{fig:prob} as a function of $T_H$ and $T_V$ for different values of $\phi$, namely $\phi=0$ (top) and $\phi=\pi$ (bottom) which correspond to the generation of $\ket{\psi^{\mp}}=\frac{1}{2}[\ket{H_AV_B}\mp\ket{V_AH_B}]$ respectively.
We can clearly see that this probability decreases faster as $T_H$ and $T_V$ deviate from the ideal values when $\phi=0$, i.e when we want to generate $\ket{\psi^-}$. This is the effect of path interference which is constructive for $\phi=\pi$ and destructive for $\phi=0$.
With the parameters adopted in our experiment we are able to generate $\ket{\psi^-}$ with probability $\mathcal{P}_{\ket{\psi^-}}=0.942$ while $\ket{\psi^+}$ is generated with probability $\mathcal{P}_{\ket{\psi^+}}=0.987$

\section*{Fabrication parameters}
\begin{figure}[t!!!!]
\centering
\includegraphics[width=\columnwidth]{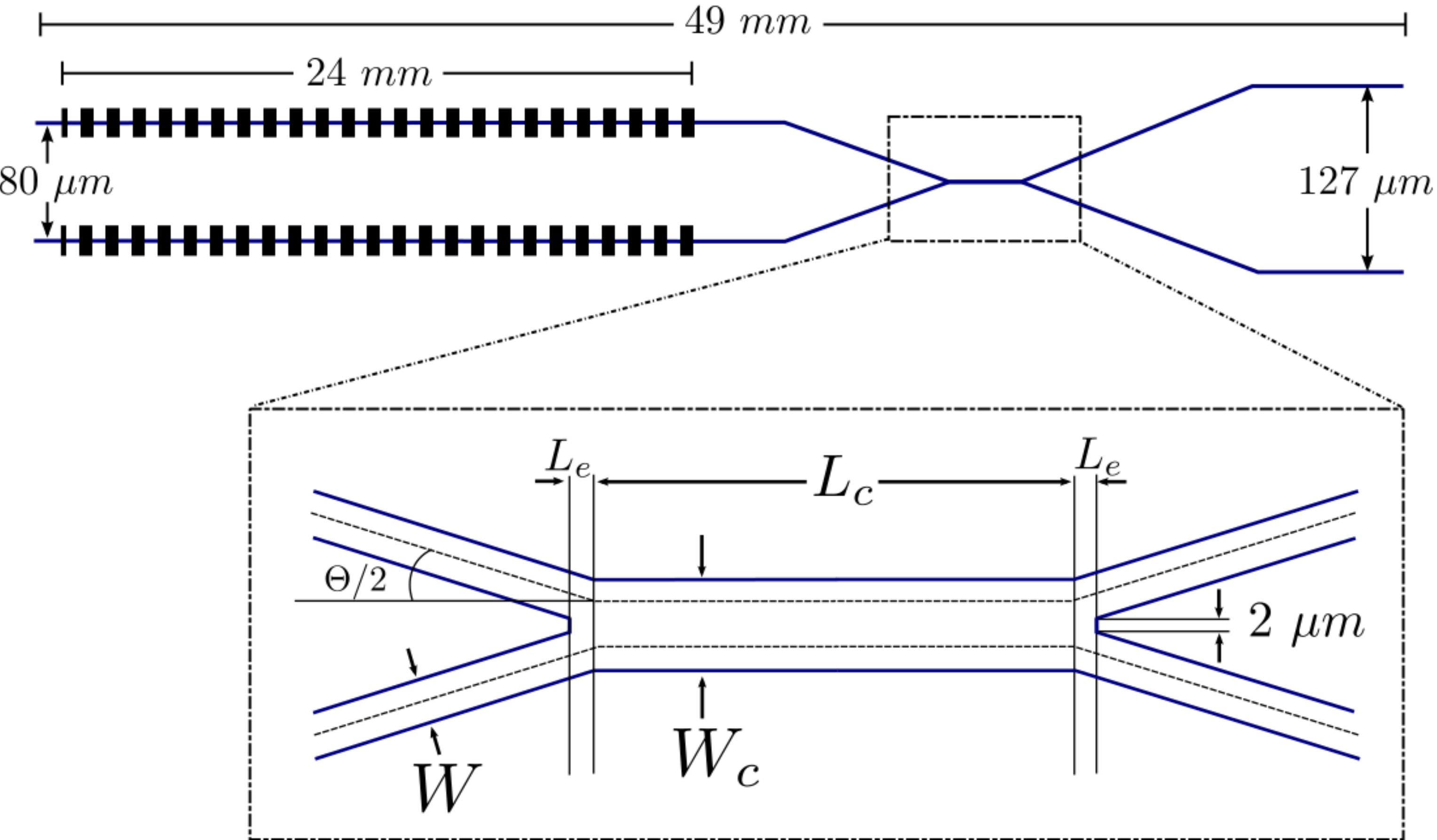}
\caption{Design of the polarization entangled source circuit. In the inset there is a zoom of the integrated polarizing beam splitter.}
\label{fig:PBS}
\end{figure}

A detailed scheme of our device is shown in Fig. \ref{fig:PBS}. The waveguide circuit is realized on lithium niobate through lithographic techniques. Titanium diffusion is carried out for about 9 hours at $1060^{\circ}C$ in oxygen  environment.
The waveguide width is $W=7\ \mu m$ and the height $85\ nm$.
The input waveguide separation is $80\ \mu m$ while the output one is $127\ \mu m$ which matches the separation of fibers in standard waveguide arrays.
The input waveguides present periodic poling with period $\Lambda=9,08\ \mu m$ for a length of $L_p=24\ mm$. These waveguides merge in the structure of a zero-gap directional
coupler used as polarization splitter. Its basic design
is shown in the inset of Fig. \ref{fig:PBS}. It consists of a central section with
length $L_c$ in which the two waveguide branches merge
into a single waveguide with width $W_c = 2W$ which is
twice the width of the single waveguides. A linear branching is performed to separate the two waveguides, i.e. the
single waveguides are inclined by an angle $\Theta$. To avoid
problems with the lithography in the range in which the
waveguide separation is smaller than $2\ \mu m$ a single tapered waveguide
with increasing width is chosen ($L_e$ region in the drawing).
In our design $W=7\ \mu m$, $L_c=480\ \mu m$ and $\Theta=0.6^{\circ}$.

\section*{Joint spectral amplitude and photon indistinguishability}
Parametric down conversion is a nonlinear process where a photon from a bright field at frequency $\omega_p$ decays into two daughter photons at lower frequencies.
\begin{figure}[b!!!!]
\centering
\includegraphics[width=0.8\columnwidth]{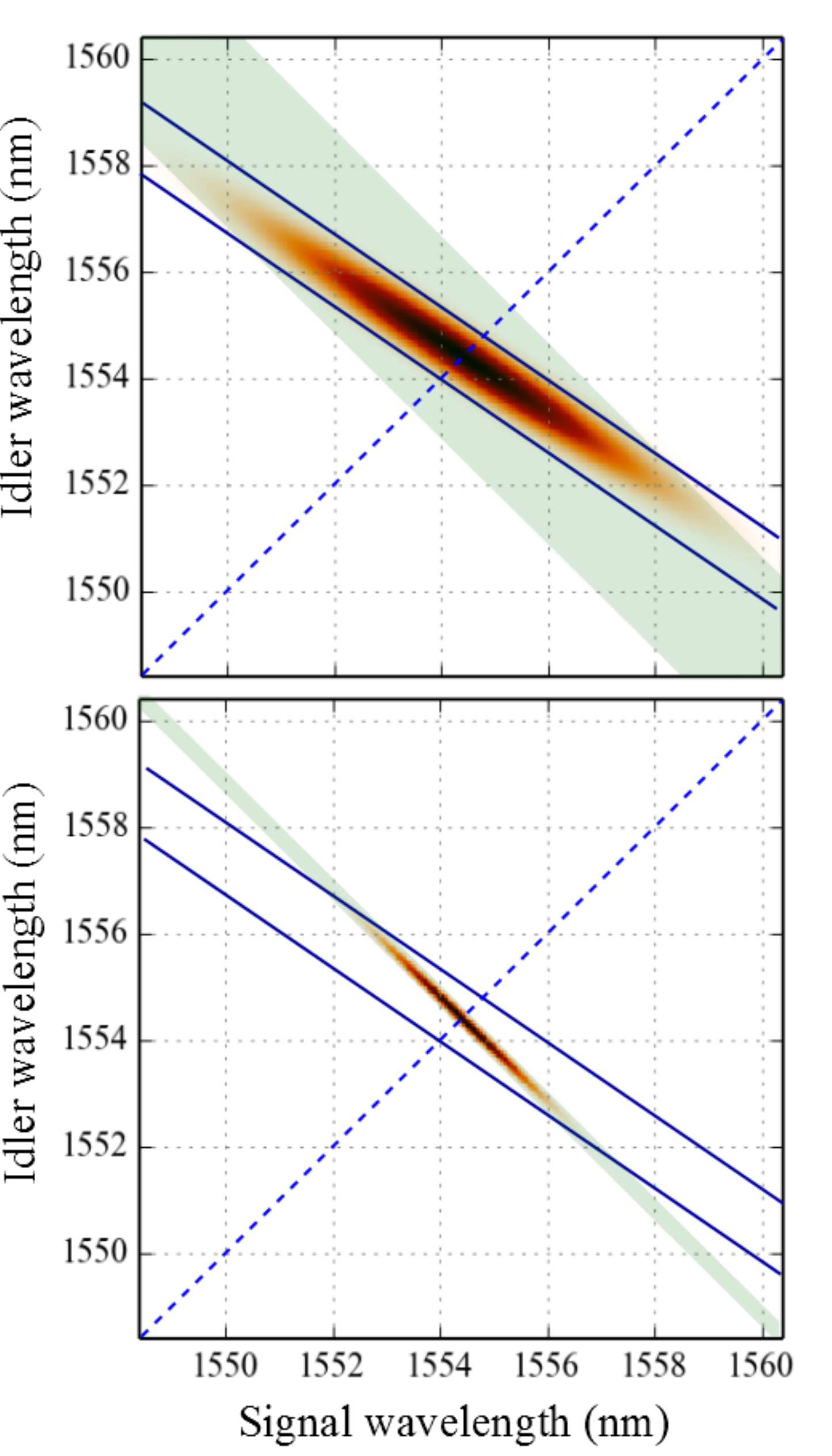}
\caption{JSA: spectrum of signal and idler generated in the PCD process. Blu continuous lines represent the PM function of type II PDC in lithium niobate, while the green shaded region is the pump spectrum where energy conservation is ensured $\omega_p=\omega_s+\omega_i$. In the case of picosecond pump pulses (top) the orientation of the JSA is dominated by the PM function while in the case of narrow band pump light (bottom) the orientation of the JSA is fixed by the pump envelope. In this second case by exchanging signal and idler -i.e. by mirroring the JSA with respect to the $45^{\circ}$ line (blue dashed)- the two JSAs completely overlap, a condition corresponding to complete spectral indistinguishability between signal and idler.}
\label{fig:JSA}
\end{figure}
For this process to occur, energy and momentum must be conserved. Momentum conservation is represented by the so called "phase matching function" which depends on the material (in our case lithium niobate) properties and the waveguide geometry. The energy conservation is ensured by the relation $\omega_p=\omega_s +\omega_i$ where the subscripts $p,s,i$ stay for pump, signal and idler respectively. On the plane $(\omega_s,\omega_i)$ this relation is represented by a $-45^{\circ}$ line whose width depends on the bandwidth of the pump spectrum (green shaded region in Fig. \ref{fig:JSA}). The final spectral properties of the PDC photons are represented on this plane by the joint spectral amplitude (JSA) which is the product of the phase matching (PM) function and the pump envelope.
For a type II degenerate process in lithium niobate for our wavelength combination the PM function is aligned at $-33.5^{\circ}$ (blue solid lines). The JSA will result in an ellipse whose orientation depends on the pump bandwidth: when the pump is broad the orientation of the JSA is dominated by the PM function (Fig. \ref{fig:JSA} top), while in the case of CW pump light -i.e almost monochromatic frequency- the orientation of the JSA is mostly determined by the pump envelope (Fig. \ref{fig:JSA} bottom).

The spectral indistinguishability of the generated photon pairs is directly connected to the orientation of the JSA. Indeed the spectral indistinguishability can be quantified by looking at the overlap between the JSA and the JSA obtained by exchanging signal and idler axes. This corresponds to mirror the JSA on the $+45^{\circ}$ line in the plane $(\omega_s,\omega_i)$ (blue dashed line in Fig. \ref{fig:JSA}) and look at the overlap between the original and the mirrored JSAs. A perfect overlap is ensured when the JSA is oriented at $-45^{\circ}$, while it decreases as the orientation deviates from this optimal value.
With picosecond pump pulses of $\approx\ 0.3\ nm$ bandwidth, as set in our experiment, the overlap is $\mathcal{O}=0.44$. In this configuration we adopted narrowband filters at the output of our device in order to eliminate the distinguishability of signal and idler spectra and consequently increase the entanglement. The overlap of the two JSAs approaches unity with CW pump light and in this configuration entanglement was generated in our device without the need of any filter.

\end{document}